\documentclass[aps,prl,twocolumn,showpacs,preprintnumbers,superscriptaddress,floatfix]{revtex4-1}

\usepackage{graphicx}
\usepackage{bm}
\usepackage{amssymb}
\usepackage{amsfonts}
\usepackage{amsmath}
\usepackage{color}

\newcommand{\comment}[1]{\textcolor{red}{COMMENT:$\ll$ #1 $\gg$}}

\newcommand{\robra}[1]{\left(#1\right)}

\newcommand{\de}[1]{d{#1}\,}

\newcommand{\be}{\begin{equation}}
\newcommand{\ee}{\end{equation}}
\newcommand{\bea}{\begin{eqnarray}}
\newcommand{\eea}{\end{eqnarray}}
\newcommand{\baa}{\begin{align}}
\newcommand{\eaa}{\end{align}}
\newcommand{\br}{{\bm r}}

\newcommand{\bk}{{\bm k}}

\newcommand{\fig}[1]{Fig.~\ref{#1}}

\newcommand{\bracket}[3]{\langle #1|#3|#2\rangle}

\newcommand{\expect}[2]{\langle {#1|#2|#1}\rangle}

\newcommand{\ket}[1]{|#1\,\rangle}

\newcommand{\inbra}[1]{{#1}}

\newcommand{\etal}{{\it et~al.}}

 \usepackage{mathrsfs} 
{\color{blue}}

\renewcommand{\comment}[1]{}

\begin{document}

\preprint{APS/123-QED} \title{Photoemission Spectroscopy and Orbital Imaging
from Koopmans-Compliant Functionals}
\author{Ngoc Linh Nguyen} \email{linh.nguyen@epfl.ch} \affiliation{Theory and
Simulations of Materials (THEOS), and National Center for Computational Design
and Discovery of Novel Materials (MARVEL), \'Ecole Polytechnique F\'ed\'erale
de Lausanne, 1015 Lausanne, Switzerland} \author{Giovanni Borghi}
\affiliation{Theory and Simulations of Materials (THEOS), and National Center
for Computational Design and Discovery of Novel Materials (MARVEL), \'Ecole
Polytechnique F\'ed\'erale de Lausanne, 1015 Lausanne, Switzerland}
\author{Andrea Ferretti} \affiliation{Centro S3, CNR--Istituto Nanoscienze,
41125 Modena, Italy} \author{Ismaila Dabo} \affiliation{Department of
Materials Science and Engineering, Materials Research Institute, and Penn State
Institutes of Energy and the Environment, The Pennsylvania State University,
University Park, PA 16802, USA} \author{Nicola \surname{Marzari}}
\affiliation{Theory and Simulations of Materials (THEOS), and National Center
for Computational Design and Discovery of Novel Materials (MARVEL), \'Ecole
Polytechnique F\'ed\'erale de Lausanne, 1015 Lausanne, Switzerland}
\date{\today}

\begin{abstract} The determination of spectral properties from first principles
can provide powerful connections between microscopic theoretical predictions
and experimental data, but requires complex electronic-structure formulations 
that fall outside the domain of applicability of common approaches, such 
as density-functional theory. We show here that Koopmans-compliant functionals, 
constructed to enforce piecewise
linearity in energy functionals with respect to fractional occupations 
--- i.e. with respect to charged excitations --- provide molecular photoemission spectra 
and momentum maps of Dyson orbitals that are in excellent agreement with experimental
ultraviolet photoemission spectroscopy and orbital tomography data.
These results highlight the role of Koopmans-compliant functionals as accurate 
and inexpensive quasiparticle approximations to the spectral potential.
\end{abstract}
\maketitle

The interpretation of experimental spectra, such as those obtained with ultraviolet photoemission spectroscopy (UPS)
or angular-resolved photoemission spectroscopy (ARPES), often requires theoretical support, due to the complexity of the data \cite{Hufner_photoelectron_1996, pendry_theory_1976}. In fact, theoretical predictions can help resolve spectral contributions coming from quasi-degenerate excitations, allowing to label each state with its native quantum numbers, or to find the correspondence between photoemission peaks and the probability density of 
the states from which electrons were emitted~\cite{stojic_ab_2008, puschnig_reconstruction_2009}. The power and accuracy of current experimental techniques, together with their microscopic resolution, strongly motivates thus the development of reliable first-principle methods able to reproduce accurately experimental spectra for different setups and photon energies, and to interpret them qualitatively and quantitatively.
From a theoretical point of view, photoemission
spectra have been studied with many-body perturbation theory 
~\cite{Onida2002, Guzzo_Gatti_Reining_photoemission_GW}, 
time-dependent extensions of density-functional theory (DFT) ~\cite{DeGiovannini2012} 
density-matrix functional theory~\cite{Sharma2013} 
or with the wave function methods of quantum chemistry~\cite{huang_study_2008, dolgounitcheva_electron_2011}.
However, due to the significant computational requirements of these approaches,
and their own limits in terms of ultimate accuracy,
applications are limited in systems' sizes and complexity~\cite{refaely-abramson_quasiparticle_2012}. 
This is the reason why simpler 
methods such as Hartree-Fock or ground state DFT are still frequently 
employed to interpret photoemission spectra \cite{morini_benchmark_2008, puschnig_reconstruction_2009}.

\begin{figure}
\includegraphics[scale=0.3]{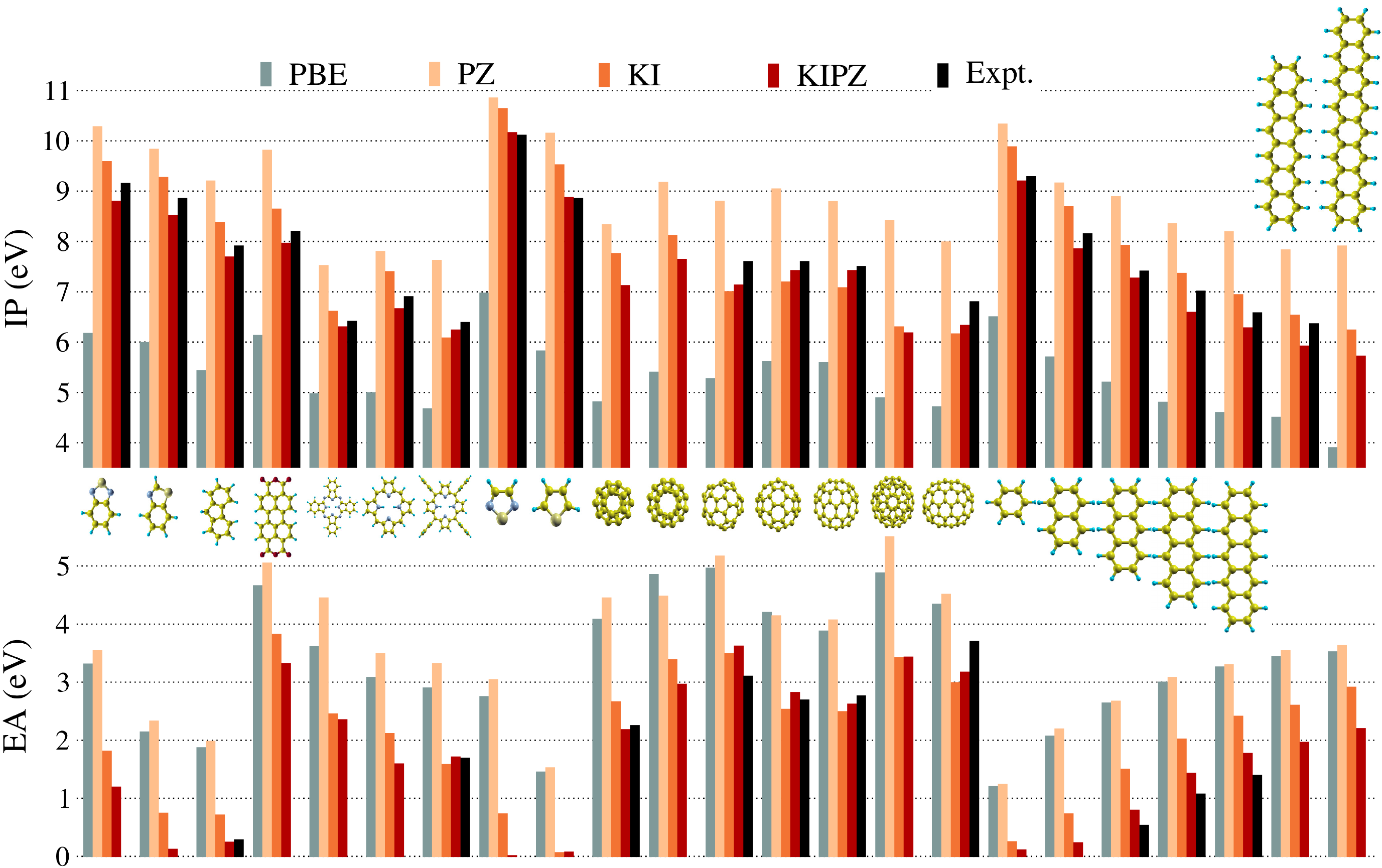}
\caption{\label{ip_ea_alls} IPs and EAs of 23 organic photovoltaic molecules, calculated using PBE or the self-interaction corrected PZ, KI, and KIPZ functionals, and
compared with available experimental data.} 
\end{figure}
\begin{table}[!htb]
\caption{\label{tab_ea_ip} Mean absolute errors (MAE) and root mean square errors (RMSE) with respect
to experiments for the IP (13 molecules, in eV) and EA (10 molecules, in eV) of a subset of molecules 
from \fig{ip_ea_alls}, for which experimental and self-consistent GW data are available 
(see Table I and II of SM~\cite{Supplemental_material}, and Refs.~\cite{blase_first-principles_2011} 
and \cite{tiago_neutral_2008}).}
\begin{ruledtabular}
\begin{tabular}{l l  r r r r r}
      & &PBE&PZ&KI&KIPZ&scf-GW\\
\hline
  IP  &MAE&2.28&1.23&0.45&0.24&0.31\\   
      &RMSE&2.33&1.25&0.47&0.28&0.34\\   
  EA  &MAE&1.57&1.72&0.54&0.25&0.27\\   
      &RMSE&1.63&1.77&0.63&0.30&0.34\\   
\end{tabular}
\end{ruledtabular}
\end{table}

\begin{figure*} \includegraphics[scale=0.6]{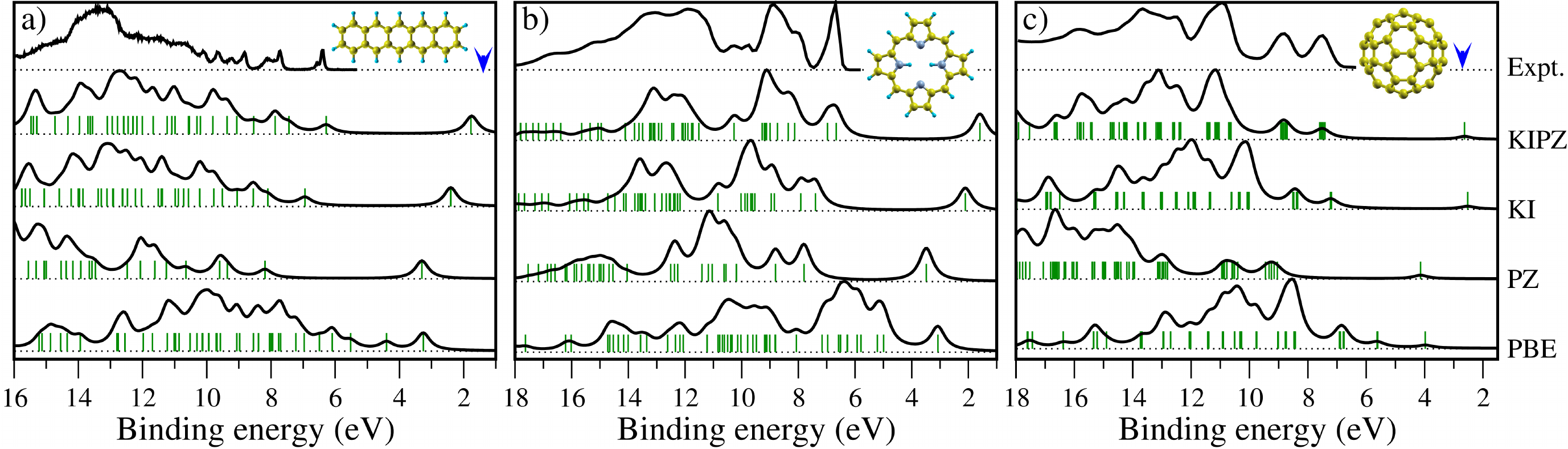}
\caption{\label{fig_3_mols} UPS spectra for (a) pentacene, (b)
porphine, and (c) fullerene C$_{60}$ calculated using the
PBE, PZ, KI, and KIPZ functionals, and plotted as a function of electron binding energy $h\nu-\hbar^2 k^2/(2m)$. 
For pentacene and porphine
the calculations are done for an incoming photon energy of 21.22 eV (corresponding to the experimental HeI radiation), and are compared 
with experimental gas-phase UPS measurements using HeI
(Ref.~[\onlinecite{coropceanu_hole_2002}] and Ref.~[\onlinecite{dupuis_very_1980}]), 
while for fullerene C$_{60}$ calculations and experimental data 
(Ref.~[\onlinecite{lichtenberger_he_1992}])
are for 50 eV photon energy.
The blue arrows mark the experimental electron affinities (corresponding to the binding energy of the lowest unoccupied molecular orbital) of pentacene and C$_{60}$, taken from 
Refs.~[\onlinecite{crocker_electron_1993}] and [\onlinecite{wang_high_1999}].}
 \end{figure*}

Recently, we introduced Koopmans-compliant (KC) functionals~\cite{Dabo2009,Dabo2010,psik_koopmans, Dabo2013, Borghi_PRB_2014} to enforce a generalized criterion of piecewise linearity with respect to the fractional 
removal or addition of an electron from any orbital [i.e. not only the highest-occupied molecular orbital (HOMO)] in 
approximate DFT functionals, and this condition is naturally akin to that of enforcing a correct description of charged excitations in photoemission experiments. Moreover, we argued~\cite{ferr+14prb} that these KC functionals
approximate directly the spectral potential, i.e. the local, frequency-dependent contraction of the electronic
self-energy that is necessary and sufficient to describe the local spectral function and the total density of 
states~\cite{ferr+14prb, Gatti2007} of a given system.

In this Letter, we illustrate the remarkable performance of this class of functionals in predicting both ultraviolet photoemission spectra and orbital-tomography momentum maps. This agreement with experiments is complemented by potential energy surfaces that preserve the quality of the base functionals
\cite{Borghi_PRB_2014} and predictions for frontier orbital energies [i.e., for ionization potentials (IPs) and electron affinities (EAs)] that are comparable and slightly superior to the state-of-the-art in many-body perturbation theory \cite{Borghi_PRB_2014}, with very moderate computational costs (no sums over empty states, cubic scaling with 
system size, and, thanks to the localization of the minimizing orbitals, directly amenable to linear-scaling 
approaches).

Photoemission spectra can be reproduced theoretically following the well established one-step model within 
the sudden approximation~\cite{pendry_theory_1976}. This approach treats the
photoexcitation as a transition from an electronic initial state
$\ket{\Phi^{\rm N}_{0}}$ -- which is the ground-state with energy $E_{\rm
0}$ -- into an excited $\rm N$-particle state $\ket{\Phi^{\rm
N}_{i,\bk}}=\ket{\Phi^{\rm N-1}_{i}}\otimes \ket{\xi_{\bk}}$ with energy $E_{i,\bk}$,
composed of the $i^{\rm th}$ excited state of the singly ionized system 
and the wave function $\xi_{\bk}$ of the ejected electron, often 
approximated by a plane-wave of wave vector $\bk$. The total photoemission intensity 
can be described, to first-order in 
perturbation theory, through Fermi's golden rule~\cite{pendry_theory_1976} as
\begin{align}\label{pes_form1} I^{(\nu)} \propto \sum_{i,\bk}|\bracket{\Phi^{\rm
N}_{0}}{\Phi^{\rm N}_{i,\bk}}{\bm{A}\cdot\bm{p}}|^2  \delta (h\nu + E_{0} - E_{i,\bk})\,,
\end{align}
which contains the squared modulus of the light-matter interaction operator in the dipole approximation -- where ${\bm A}$ is the
semi-classical electromagnetic field and $\bm{p}$ is the momentum operator 
of the electron -- and a $\delta$ function for energy conservation, requiring the 
incoming photon energy $h\nu$ to be equal to the total excitation energy $E_{i,\bk}-E_{0}$.
Equation~\eqref{pes_form1} can be written in terms of single-particle Dyson orbitals, $\phi^{\rm d}_i(\br) = \bracket{\Phi^{\rm N-1}_{i}}{\Phi^{\rm N}_0}{\hat{\Psi}(\br)}$
($\hat{\Psi}(\br)$ being the annihilation operator for an electron at point $\br$
in space)~\cite{walter_photoelectron_2008}, as
\begin{align}\label{pes_form3} 
I^{(\nu)} \propto \sum_{i,\bk} |\bracket{\phi^{\rm
d}_{i}}{\xi_{\bk}}{ \bm{A}\cdot \bm{p}}|^2  \delta \robra{h\nu - \varepsilon^{\rm d}_i -
\frac{\hbar^2 \bk^2}{2m}}\,, 
\end{align}
where now the excitation energy is expressed in terms of the kinetic energy 
$\hbar^2 \bk^2/(2m)$ of the ejected electron and its binding energy, i.e., the 
difference between the energy of the $i^{\rm th}$ excited state of the 
ionized system and the ground state energy of the neutral system. 
The usual way to present a photoemission spectrum is to plot, for a given $h\nu$, the number of photoelectron counts as a function of $h\nu-\hbar^2 \bk^2/(2m)$, which is the definition of binding energy of photoelectron in terms of the observables of the photoemission process.

From a theoretical point of view, the electron binding energy is equal to minus the pole $\varepsilon^{\rm d}_i$ of the one-body Green's function.
Dyson orbitals, together with their binding energies $\varepsilon^{\rm d}_i$, 
should therefore in principle be determined by solving 
quasiparticle equations within the framework of many-body 
perturbation methods \cite{Guzzo_Gatti_Reining_photoemission_GW, barbiellini_dyson_2004}.
Due to the complexity of these approaches, common practice still relies 
on exploiting molecular orbitals computed with e.g.
Kohn-Sham (KS) DFT~\cite{Chong2002, vogel_photoionization_2011}, 
where $\{\phi^{\rm d}_{i}\}$ and their corresponding $\{\varepsilon^{\rm d}_i\}$ are
approximated by the single-particle eigenstates $\{\varphi_{i}\}$ and
eigenvalues $\{\varepsilon_i\}$, respectively (more details on the calculation of $I^{(\nu)}$ can be found in the Supplemental Material (SM)~\cite{Supplemental_material}).  
While it has been argued~\cite{Duffy1994, Chong2002} that the exact KS
eigenstates and eigenvalues are an accurate approximation of Dyson orbitals and quasiparticle excitations, particularly in an energy window close to the
HOMO energy, approximate density functionals such as the local density
approximation (LDA) and the generalized gradient approximations (GGAs) are known to produce electronic eigenvalues which are in poor correspondence with
physical particle removal energies, making their predictions for photoemission spectra unreliable. 
Even HOMO eigenvalues, that in exact KS DFT correctly reproduce the negative of the IP, are strongly under-estimated due to the well-known self-interaction error intrinsic to LDA and GGA functionals~\cite{Perdew1981,cohe+08sci,psik_koopmans} (see the results of the Perdew-Burke-Ernzerhof (PBE) functional~\cite{perdew_generalized_1996} in \fig{ip_ea_alls} and Table \ref{tab_ea_ip} for 23 molecules relevant for photovoltaic applications), which is also responsible for the spatial over-delocalization of the total charge density and of electronic wave
functions~\cite{cohen_challenges_2011,cohe+08sci}. While a number of methods have been proposed to correct for self-interaction,
such as the Perdew and Zunger correction
(PZ)~\cite{Perdew1981}, DFT+Hubbard U to reduce
hybridization and delocalization of $d$ or $f-$orbitals~\cite{Cococcioni2005, kuli+06prl}, and range-separated hybrid
functionals~\cite{refaely-abramson_quasiparticle_2012}, in this Letter we
take the view of Ref.~[\onlinecite{ferr+14prb}] that KC functionals are beyond-DFT approaches
which approximate the exact spectral potential, thus recovering reliable particle removal energies, while preserving the accuracy of the underlying energy functional relative to total energy and density.

As a reminder, KC functionals are obtained by removing,
orbital-by-orbital, the non-linear Slater contribution as
a function of fractional occupation from the total energy
functional, and adding in its lieu a linear (Koopmans)
term in the occupations. We focus here on the KI and KIPZ functionals, which can be obtained by correcting any approximate functional $E^{\inbra{{\rm app}}}$ as $E^{\inbra{{\rm KC}}} = E^{\inbra{{\rm app}}} +\alpha \sum_{i} \Pi_i^{\rm KC}$, where for KI, $\Pi_i^{\rm KC}$ is
\begin{align}\label{Eq:KC_KI} \Pi^{\rm KI}(\rho_{i}) = &-\int_0^{f_i}{\expect{\phi_i}{H^{\rm app}(s)} \de{s}} \nonumber \\
                                                        &+ f_i \int_0^1{\expect{\phi_i}{H^{\rm app}(s)} \de{s} } \,
\end{align}
and for KIPZ
\begin{align}\label{Eq:KC_KIPZ} &\Pi^{\rm KIPZ}(\rho_{i}) = \Pi^{\rm KI}(\rho_{i}) -f_i E_{\rm Hxc} [|\phi_i|^2].
\end{align}
In the above equations, $\rho_i = f_i |\phi_i|^2$, and $H^{\rm app}(s)$ is the approximate KS Hamiltonian calculated with orbital $\phi_i$ fractionally occupied $s$ (we refer to Ref.~[\onlinecite{Borghi_PRB_2014}] for details$-$the expressions above being equivalent to Eqs.~(25) and (A6), (27) and (A13) of Ref.~\cite{Borghi_PRB_2014}, respectively). We note that the multiplicative factor $\alpha\in [0,1]$ in the definition of $E^{\rm KC}$ acts as a simplified electronic screening function, and it is chosen so that IP of the system under consideration is equal to EA of the same system, deprived of an electron. As shown by Dabo $\etal$~\cite{Dabo2010, Dabo2013}, and Borghi
$\etal$~\cite{Borghi_PRB_2014}, this constant screening is sufficient to
accurately predict IP and EA energies from the eigenvalue spectrum of a variety of molecular systems, although a more sophisticated orbital-dependent choice might be convenient in the case of more complex or extended systems. 

Koopmans' orbital-by-orbital linearity condition is more stringent than the
piecewise linearity condition satisfied by the exact KS DFT ground-state
energy as a function of fractional changes in the occupation of the HOMO, and in
turn it provides a more general orbital-dependent framework.
In fact, the derivatives of Koopmans' $\Pi_i^{\rm KC}$ corrections with respect to orbital 
densities can be interpreted as simplified local self-energies
$\hat{\Sigma}^{(i)}(\br)=\frac{\delta \Pi^{\rm KC}(\rho_{i})}{\delta \rho_{i}(\br)}$
modifying the KS Hamiltonian of the approximate functional \cite{ferr+14prb}
\begin{align}\label{Eq:KC_h} H^{\rm KC}_{mn} = H^{\rm app}_{mn} +
\alpha \sum_{i}{\Sigma^{(i)}_{mn}}\,\, ; \end{align} 
in particular, in the basis of the {\it variational} (and localized) orbitals $\{\phi_i\}$ that minimize $E^{\rm KC}$ we have
\begin{align}\label{Eq:KC_Sigma}
\quad \Sigma^{(i)}_{mn}=\delta_{im}\bracket{\phi_{m}}{\phi_{n}}{\hat{\Sigma}^{(i)}}\,.  
\end{align}
Definitions Eqs.~\eqref{Eq:KC_KI} to~\eqref{Eq:KC_Sigma} are valid both for orbitals
belonging to the valence manifold, and for empty orbitals, for which
$\rho_{i}$ is the orbital density of a virtual electron added to or removed from the system. Within this approach, which is rigorously applicable to gapped systems,
both finite and extended, the Hamiltonians for valence and empty states are
decoupled, i.e., any matrix elements between filled and empty states introduced
by Koopmans' corrections are projected out.
Diagonalization of the modified Hamiltonian Eq.~\eqref{Eq:KC_h}, whose orbital-density dependent
antihermitian part vanishes at the energy
minimum~\cite{pede+84jcp,sten-spal08prb}, leads to generalized eigenvalues $\{\varepsilon_i\}$ and {\it canonical} eigenvectors $\{\varphi_{i}\}$ \cite{Borghi_PRB_2014}. The former can be interpreted as particle removal (if belonging to filled states) or addition (if belonging to empty states) energies, while the latter have an interpretation in terms of Dyson orbitals~\cite{ferr+14prb}. 
The self-energy operator $\hat{\Sigma}^{(i)}(\br)$ can either be constant, as within the KI functional, or local in space
as in KIPZ; in this case it modifies not only the eigenvalue spectrum of the system, but also its ground-state energy and density.

In \fig{ip_ea_alls} and Table \ref{tab_ea_ip}, we highlight the accuracy of KC functionals in predicting the energy of frontier orbitals by comparing IPs and EAs for a set of 23 molecules relevant for photovoltaic applications, for which KI and KIPZ show a performance which is comparable in quality to that of many-body perturbation methods (self-consistent GW). We next show (\fig{fig_3_mols}) a few theoretical photoemission spectra, confirming how KC functionals can successfully predict also the binding energies of deeper states. 
The three panels of~\fig{fig_3_mols} display indeed a remarkable agreement between the simulated KI and KIPZ spectra and experimental spectra (results for all 23 molecules of~\fig{ip_ea_alls} at different photon energies are shown in the SM~\cite{Supplemental_material}). The improvement over PBE predictions is evident not only in the peak positions, but also for shapes and intensities. On the contrary, PZ over-estimates peak positions, with an opposite bias with respect to PBE.
All orbital-density-dependent functional minimizations in this work are performed using a modified version of the Car-Parrinello molecular dynamics code of the Quantum-\textsc{espresso} distribution~\cite{Giannozzi2009}, with {\it ad hoc} post-processing developed for the calculation of photoemission spectra. 

\begin{figure} \includegraphics[scale=0.3]{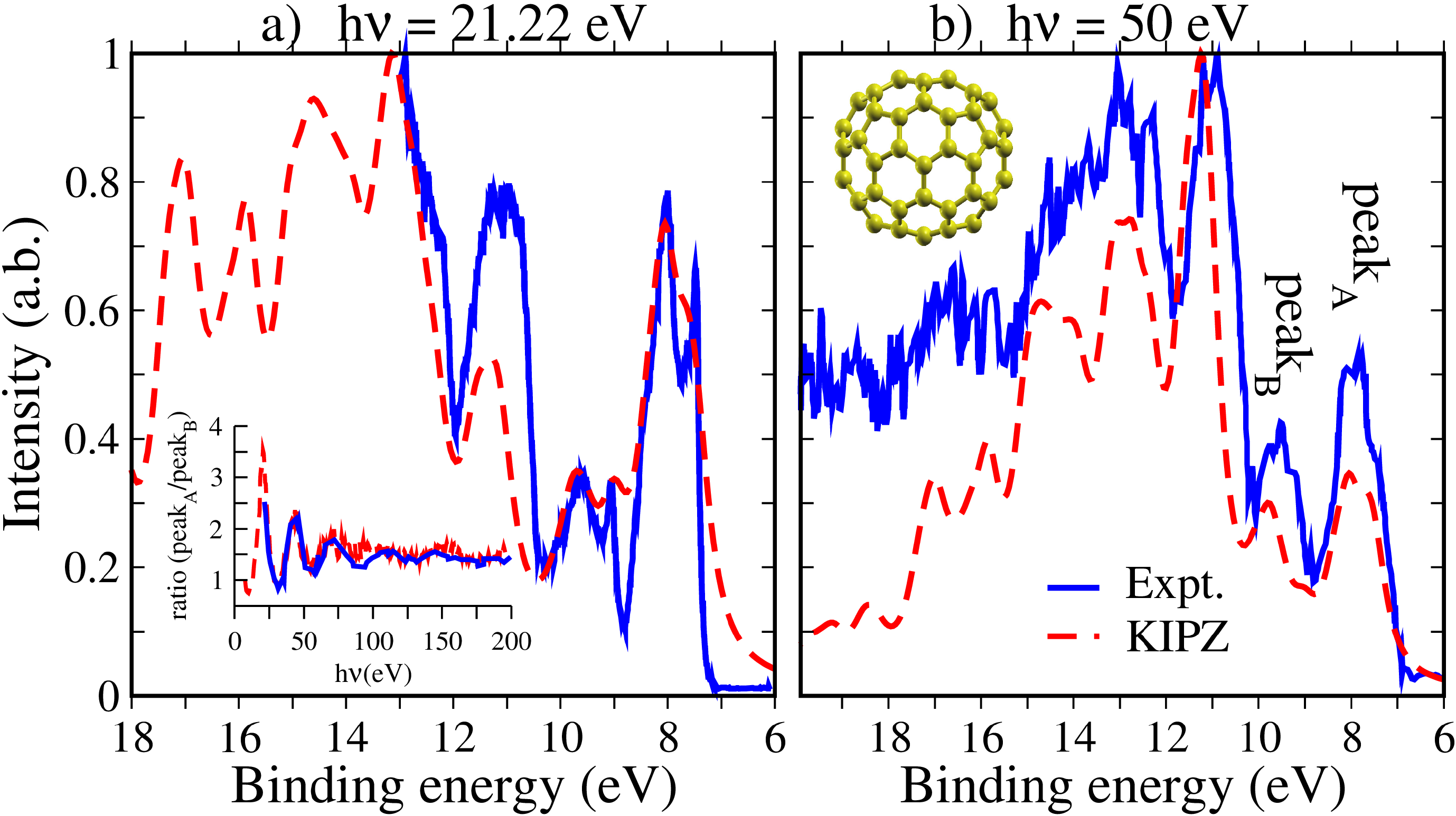}
\caption{\label{ups_vs_hnu1} Theoretical predictions for the photoemission spectrum of the fullerene C$_{70}$,
performed at incoming photon energies of 21.22 eV and 50 eV. They are compared
with experimental gas-phase photoemission data at the same photon energies, taken from
Refs.~[\onlinecite{lichtenberger_he_1992}] and~[\onlinecite{korica_2010}],
respectively. The inset shows the photoemission intensity ratio of the A and B
peaks computed at different photon energies, using the KIPZ functional (red
dashed line); the experimental data (blue line) are taken from
Ref.~[\onlinecite{korica_2010}].} \end{figure}
%
We believe there are two main explanations for the success of the KI and KIPZ functionals: (i) KI corrects the KS eigenvalues of approximate DFT by aligning them to particle removal energies through Koopmans' condition, while (ii) KIPZ adds to this feature the exactness in the one-electron limit, in which it recovers the Rydberg series of the hydrogen atom. This latter property (i.e., recovering the $1/r$ behavior of the exact KS potential) is essential in the development of novel functionals and plays an important role in the prediction of fundamental gaps and 
excitation energies~\cite{refaely-abramson_quasiparticle_2012}. At variance with the KI functional, the KIPZ functional is able to modify not only the electronic excitation energies of approximate DFT, but also the manifold of electronic orbitals (i.e. the single-particle KS density-matrix)~\cite{Borghi_PRB_2014}. A change in the density-matrix has repercussions on the values of both photoemission peak 
intensity and position of every electronic excitation, which explains why the KIPZ functional yields a more accurate description of experimental data.
Our results prove that KC functionals are successful also in capturing the change in photoemission peak positions and intensities when changing the energy of the incoming photon. \fig{ups_vs_hnu1} displays for instance the results for fullerene C$_{70}$, compared with experiment, for incident photon energies of 21.22~eV
(Ref.~[\onlinecite{lichtenberger_he_1992}]) and 50~eV
(Ref.~[\onlinecite{korica_2010}]), respectively. 
For both photon energies, the agreement between theoretical and experimental spectra for peak positions and relative intensities is very good. Experiments indicate also that the first three peaks of the spectrum of C$_{70}$ are subject to more pronounced variations than the others when changing the energy of the incident photon $h\nu$. In particular, the relative intensities of the first two peaks are characterized by oscillations (see Ref.~\onlinecite{korica_2010} and references therein) that our photoemission plane-wave model, despite being quite simple, can describe accurately (see inset of~\fig{ups_vs_hnu1}), without resorting to the more sophisticated spherical shell model~\cite{xu_oscillations_1996}.
\begin{figure} \includegraphics[scale=0.3]{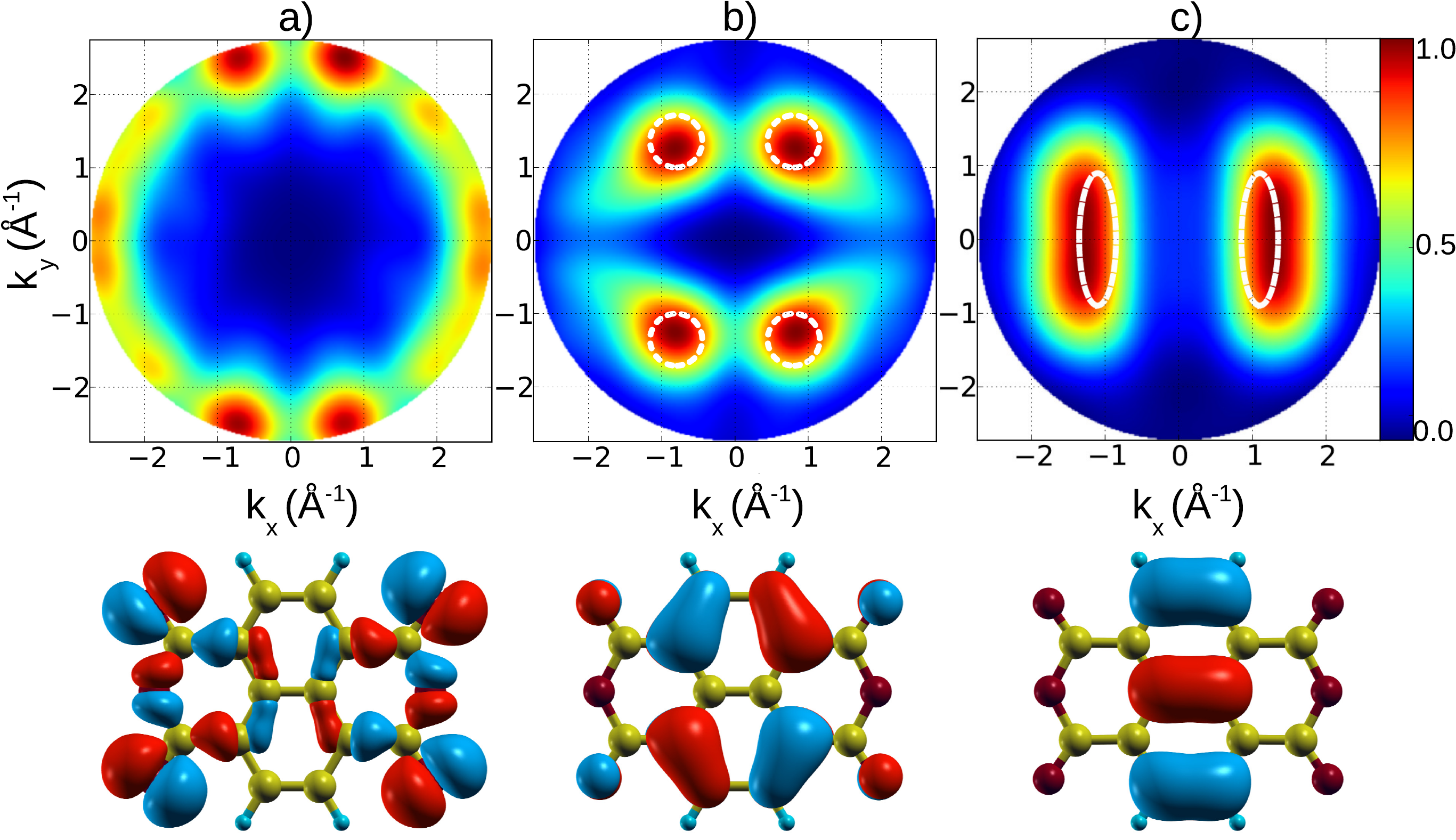}
\caption{\label{mo_arpes} Square Fourier transforms of different NTCDA
molecular orbitals computed with different methods: HOMO of PBE (top left), HOMO
of KIPZ (top center) and HOMO-1 of KIPZ (top right). The white dashed circles in
the $(k_x,k_y)$-momentum maps represent experimental intensity iso-lines taken from
Ref.~[\onlinecite{dauth_orbital_2011}]. The bottom frame shows the density
isosurfaces of each corresponding orbital.} \end{figure}

In the last part of this Letter, we discuss the ability of KC functionals to predict data from orbital tomography. This technique consists in exploiting angle-resolved photoemission spectroscopy to extract momentum maps of molecular orbitals~\cite{puschnig_reconstruction_2009, dauth_orbital_2011}. Due to the complexity of spectroscopic data, the deconvolution of orbital maps from photoemission results requires the support of theoretical UPS simulations, usually performed starting from the eigenstates of KS DFT with the PBE approximation for exchange and correlation energies~\cite{puschnig_reconstruction_2009}. Unfortunately, the accuracy of PBE is particularly compromised in systems whose ground-state wave function is composed of KS eigenstates with very different spatial character, e.g., localized vs delocalized. This difference in localization is such that the eigenvalues corresponding to these eigenstates carry unequal self-interaction biases, so that the overall orbital ordering disagrees with experiment.
An example of this flaw can be found in 1,4,5,8-naphthalene-tetracarboxylic dianhydride (NTCDA) in which localized (on the anhydride side groups) and delocalized (on the naphthalene core) outer valence orbitals coexist for very close orbital energies~\cite{korzdorfer_when_2009, korzdorfer_single-particle_2010}. In
NTCDA, their energy difference is smaller than the self-interaction biases on two
of the lowest-lying localized orbitals, so that these are pushed higher in
energy -- up to HOMO and the second highest occupied molecular orbital (HOMO-1) positions. This contradicts experimental results
[see Fig.~\ref{mo_arpes}(a)] and recent calculations combining ARPES data of
NTCDA molecules adsorbed on a Ag(110) surface and a
generalized optimized effective potential approach, with which Dauth {\it
et al.}~\cite{dauth_orbital_2011} demonstrated that HOMO and HOMO-1  states
should correspond to two delocalized orbitals with an experimental energy
difference of about $\Delta E = 0.44$ eV (measured from the kinetic energy
difference of photoelectrons ejected from either state). We
find that the KIPZ functional not only predicts an ordering of molecular
orbitals which agrees with experiment (see Fig.~\ref{mo_arpes}b and~\ref{mo_arpes}c ), but also an
energy difference between HOMO and HOMO-1 orbitals -- of about $\Delta E=0.41$
eV -- reproducing very closely the experimental difference than other approaches,
such as PZ ($\Delta E=0.17$ eV) and KI ($\Delta E=0.12$ eV). 

In conclusion, we find that KC functionals, and in particular KIPZ, can be reliable and accurate theoretical tools for the prediction of UPS spectra, and can also be exploited successfully in constructing orbital tomography momentum maps, in close agreement with ARPES measurements. As argued in Ref.~[\onlinecite{ferr+14prb}], these functionals provide a beyond-DFT approach where the spectral potential, rather than the exchange-correlation one, is directly approximated.  


\bibliography{biblio} \bibliographystyle{aip}
\end{document}